\documentclass[comsoc]{IEEEtran}
\usepackage{cuted}
\usepackage{amsmath,amssymb,epsfig,cite,footnote,authblk, epstopdf}
\usepackage{listings, color, graphicx, xcolor}
\usepackage{wrapfig}
\usepackage{bbm}
\DeclareFixedFont{\ttb}{T1}{txtt}{bx}{n}{10} % for bold
\DeclareFixedFont{\ttm}{T1}{txtt}{m}{n}{10}  % for normal
\definecolor{deepblue}{rgb}{0,0,0.5}
\definecolor{deepred}{rgb}{0.6,0,0}
\definecolor{deepgreen}{rgb}{0,0.5,0}
\definecolor{backcolour}{rgb}{0.95,0.95,0.92}
\definecolor{codegrey}{rgb}{0.5,0.5,0.5}
\newcommand\pythonstyle{\lstset{
language=Python,
basicstyle=\ttm,
tabsize=2,
backgroundcolor=\color{backcolour},
otherkeywords={self,__future__,as,linspace},             % Add keywords here
keywordstyle=\ttb\color{deepblue},
emph={MyClass,__init__,True,False,None},          % Custom highlighting
emphstyle=\ttb\color{deepred},    % Custom highlighting style
stringstyle=\color{deepgreen},
numberstyle=\tiny\color{codegrey},
numbers=left,                    
numbersep=5pt, 
stepnumber=1,                    % the step between two line-numbers. If it's 1, each line will be numbered
showtabs=false,
commentstyle=\color{magenta},
frame=tb,                         % Any extra options here
showstringspaces=false            %
}}

\lstnewenvironment{python}[1][]
{
\pythonstyle
\lstset{#1}
}
{}

\newcommand\pythonstylepseudo{\lstset{
language=Python,
basicstyle=\ttm,
tabsize=2,
backgroundcolor=\color{backcolour},
otherkeywords={self},             % Add keywords here
numberstyle=\tiny\color{cyan},
numbers=left,                    
numbersep=5pt, 
stepnumber=1,                    % the step between two line-numbers. If it's 1, each line will be numbered
showtabs=false,
}}

\lstnewenvironment{pseudopython}[1][]
{
\pythonstylepseudo
\lstset{#1}
}
{}

\usepackage{color}
\usepackage{hyperref}
\hypersetup{
    colorlinks,%
    citecolor=blue,%
    filecolor=blue,%
    linkcolor=blue,%
    urlcolor=blue
}
\definecolor{bluecolor}{rgb}{0,0.,1.}

\definecolor{redcolor}{rgb}{.7,0.,0.}

\newcommand{\es}[1]{\begin{equation}\begin{split}#1\end{split}\end{equation}}
\newcommand{\est}[1]{\begin{equation*}\begin{split}#1\end{split}\end{equation*}}

\newcommand{\R}{\mathbb{R}}

\newcommand{\h}{\mathcal{H}}

\newcommand{\by}{\boldsymbol{y}}
\newcommand{\bX}{\boldsymbol{X}}
\newcommand{\dd}{\textrm{d}}

\newcommand{\bxi}{\boldsymbol{\xi}}

\begin{document}
\title{Temporal connectivity in finite networks with non-uniform measures}
\author[1]{Pete Pratt}
\author[1]{Carl P. Dettmann}
\author[2]{Woon Hau Chin}
\affil[1]{School of Mathematics, University of Bristol, University Walk, Bristol, BS8 1TW, UK}
%\affil[2]{School of Mathematics, University of Bristol, University Walk, Bristol, BS8 1TW, UK}
\affil[2]{Toshiba Telecommunications Research Laboratory, 32 Queens Square, Bristol, BS1 4ND, UK}
\maketitle
\pagenumbering{arabic}

\textit{Abstract}

Soft Random Geometric Graphs (SRGGs) have been widely applied to various models including those of wireless sensor, communication, social and  neural  networks.
SRGGs are constructed by randomly placing nodes in some space and making pairwise links probabilistically using a connection function that is system specific and usually decays with distance. 
In this paper we focus on the application of SRGGs to wireless communication networks where information is relayed in a multi hop fashion, although the analysis is more general and can be applied elsewhere by using different distributions of nodes and/or connection functions.
We adopt a general non-uniform density which can model the stationary distribution of different mobility models, with the interesting case being when the density goes to zero along the boundaries. 
The global connectivity properties of these non-uniform networks are likely to be determined by highly isolated nodes, where isolation can be caused by the spatial distribution or the local geometry (boundaries).
We  extend the analysis to temporal-spatial networks where we fix the underlying non-uniform distribution of points and the dynamics are caused by the temporal variations in the link set, and explore the probability a node near the corner is isolated at time T. 
This work allows for insight into how non-uniformity (caused by mobility) and boundaries impact the connectivity features of temporal-spatial networks. 
We provide a simple method for approximating these probabilities for a range of different connection functions  and  verify them against simulations.
Boundary nodes are numerically shown to dominate the connectivity properties of these finite networks with non-uniform measure.

\section{Introduction}
Since the seminal paper by Gilbert in 1961 \cite{gilbert1961random} where Random Geometric Graphs (RGGs) (originally labelled random plane networks) were first introduced, they have been applied to model the spread of diseases, fires and information; in fact even in the original paper wireless networks were proposed as a relevant application.
More recently, RGGs have been used to help model how devices are deployed, and subsequently interact, in wireless sensor networks \cite{kenniche2010random}, for example with a mesh architecture where there is no fixed infrastructure and information is transferred in a multi-hop fashion. 
In the classical RGG points are randomly placed in some space and two points form a link if their distance is less than some critical distance $r_0$ \cite{dettmann2018confined, walters2011random}.

The original RGG model was extended by Waxman in 1988 \cite{waxman1988routing},focusing on packet routing in wireless networks to include probabilistic connections, more recently coined Soft Random Geometric Graphs (SRGGs) \cite{Penrose16,Krioukov16,MP15}.  
The additional source of randomness produced by the probabilistic connection functions generates a wider array of  applications including neural and social networks  \cite{roudi2004associative, shiino1992self, wong2006spatial, cho2011friendship} and a wider range of communication networks.

Even for a spatial network where the node locations are fixed, the set of edges can vary with time due to random link failures \cite{kar2008sensor } which themselves can often be spatially correlated \cite{kar2010distributed}.
For example in a wireless sensor network where the location of nodes remains unchanged, a node may go from being connected to disconnected in consecutive time slots  due to fluctuations in the communication channel.
Thus, it makes sense to talk about both the temporal and spatial features of these networks, herein referred to as temporal spatial networks. 
The impact of these random failures can be mitigated in a very mobile environment as problems of a node being located form a neighbour is very short lived\cite{gupta2000capacity, grossglauser2001mobility}; equivalently one can think of the time needed to transmit information, and the time for a node to change its location as having a similar time scale. 
However, in reality retransmissions in smart devices occur on a much smaller time scale compared with human mobility, say; so to  accommodate these different time scales we assume the locations of nodes are fixed throughout time, but connections are made during each time step according to a connection function $\h$ independent of the past.
In particular we focus on the local property of node isolation near a corner where the distribution of points is non-uniform (and can go to zero at the boundary) over multiple time slots and explore how these local properties impact the global picture of connectivity.

The connectivity of SRGGs is closely related to that of  continuum percolation where, unlike it's classical counterpart,  the locations of nodes in the graph are random. 
Early bounds were given in Ref\cite{gilbert1961random} on the conditions needed for there to exist a connected component of infinite size (giant component) in a RGG in $\R^2$  by relating the problem to a branching process (lower bound) and bond percolation on the square lattice (upper bound).
For a fixed $r_0$ there exists a percolation transition where a node goes from belonging to a component of finite size almost surely, to being connected to the giant component with positive probability. 
Various work has focused on improving these bounds discussed in Ref\cite{walters2011random}, while others have looked at different regimes, for example in sparse communication networks no additional infrastructure is required in two dimensions when devices are well scattered, this is true in one dimension  \cite{dousse2002connectivity}. 
More recent work has focused on local power management (vary the connection range $r_0$) to achieve connectivity \cite{glauche2003continuum, iyer2012nonuniform, gouere2008subcritical}, a result extended to Poisson hole networks, (the holes represent regions nodes cannot be, see Ref\cite{lee2012interference}), which model competing cognitive radio networks  \cite{yemini2016simultaneous, sarkar2017continuum}.
Percolation on SRGGs is less well studied, with some of the more notable work being done on networks where interference is included so a link between any two nodes also depends on the location of other nodes in the network \cite{dousse2006percolation, Vaze12}.

For networks in a finite domain a more natural 
and stronger condition than that of percolation
is one of full connectivity, $P_{fc}$, i.e. when is there a multi-hop path between any two nodes in the network. 
Understanding the bottle neck to $P_{fc}$ is of great importance in applications of wireless mesh networks, for example where disconnected nodes may represent isolated sensors which hold important information or else dissatisfied customers.
In the classical RGG the transition from disconnected to fully connected occurs when there are no more isolated nodes\cite{penrose1997longest}, which are located far from the boundary; a result which was later extended to SRGGs \cite{mao2013connectivity}.
Interestingly, this work highlights how the local effects of isolation probabilities determine the macroscopic behaviour of $P_{fc}$ in the limit as the number of nodes tends to infinity and the typical connection range goes to zero.
Similar work has been done on the RGG with a large class of densities in 2 dimensions by Hsing and Rootzen \cite{HR05}, in higher dimensions when nodes are normally distributed \cite{penrose1998extremes} and when the connection range is location dependent \cite{iyer2012nonuniform}. 

In finite networks it is likely that border effects will dominate. 
A cluster expansion approach was utilised by Coon, Georgiou and one of the present authors to show that $P_{fc}$ can be decomposed into contributions from the bulk and the different types of boundary, where the latter tend to dominate \cite{coon2012full}.  
This result was extended to a more general class of connection functions showing that boundaries can obstruct $P_{fc}$ in dense networks \cite{dettmann2016random}.

A feature of wireless mesh networks is that they have no fixed infrastructure as the locations of nodes may vary with time as they move according to some mobility model. 
Simply put, mobility models are a set of rules (usually probabilistic in nature) that describe the movement of nodes. 
The complexity of the mobility model is inversely correlated to its mathematical tractability.
For example, one of the simplest mobility models is a Random Walk (RW), or Brownian motion, where a new direction of travel is chosen at random at each time step, with trajectories of paths being reflected off any boundaries \cite{camp2002survey}. 
The RW is recurrent in dimensions $\le 2$ meaning that a single node explores the whole of the domain \cite{bandyopadhyay2007stochastic},  consequently,  a uniform spatial distribution can be used to approximate the mobility of dense mesh network in this case \cite{gong2014interference}.

However, the spatial distribution of nodes is unlikely to be uniform as people tend to congregate around popular places such as city centres and this behaviour can be captured by the stationary distribution of the Random Waypoint Mobility (RWP)  Model \cite{bettstetter2003node}.
In the RWP model each node moves independently form one another, so it suffices to describe the process of a single node. 
A single node is placed in the domain uniformly at random, chooses a waypoint uniformly at random  and travels toward it in a straight line with a speed taken also from a uniform distribution. 
Once at the destination, the node pauses for some time, taken from some appropriate distribution,  with probability $p_T$ and then selects its next waypoint, independently from the past. 
If $p_T = 0$ then the density goes to zero along the boundary. The RWP converges to a stationary distribution, with the majority of nodes found within the bulk due to the travelling paths \cite{bettstetter2003node}.

Networks with other non-uniform measures have been studied, with more recent work focusing on their fractal nature \cite{chen2017capacity, chen2017capacity2} where it was shown that the approximation of isolated nodes causing disconnectivity improves in this case \cite{dettmann2017isolation}. 
This in essence suggests isolated nodes in networks with non-uniform measures are "more isolated" than their counterparts in uniform networks.

To date there has been little focus on temporal-spatial networks where the dynamics on the network are caused by the probabilistic nature of links, node mobility or both.
One approach is to assume the nodes have infinite mobility resulting in no spatial correlation between time slots, or alternatively fix the underlying distribution of nodes, either way this has largely been focused on the uniform case\cite{haenggi2013local, dettmann2017isolation}.
When the node locations are fixed, uniformly distributed on the torus (mitigating edge effects by using periodic boundary conditions) and links are drawn during each time slot, connectivity is determined by those nodes which are "highly isolated"   \cite{dettmann2017isolation}.
When the nodes are mobile, and follow a RW in $\R^d$, Ref \cite{peres2013mobile} obtained asymptotic results for how long a node takes to connect to any other node in the graph when the connection model is that of the RGG. 

In this work we address the question of  how boundary and non-uniform densities impact the local and global connectivity properties of temporal spatial networks.
Of particular interest in this paper are wireless communication networks, where the random location of points represent mobile smart devices, and the connection functions represent different channel conditions. 
For example, a wireless sensor network is likely to have close range connections due to power constraints so will be closer to the classical RGG, where as communication networks will have longer ("softer") connections derived from an information theoretic standpoint.
The impact of human mobility is approximated by a fixed non-uniform distribution of users, where we assume that the time scale for transmissions is much smaller than that of human mobility. 
An interesting example, which is so far unexplored in the context of full connectivity  with the exception of Ref \cite{dettmann2017isolation}, is when the density  goes to zero along the boundaries, with a motivating example being  the stationary distribution of the RWP with no pause time. 

Although the connection functions are motivated from a wireless networks perspective our analysis is general enough to incorporate connection functions from other literature. 
Finally, we make comparisons between how long one node near the corner is isolated for  compared with how long any node in the network is isolated for which provides an approximation of $P_{fc}$ in temporal spatial networks.

The paper is structured accordingly: in Sec \ref{s:model} we define the model and introduce the tools required for the analysis; Sec \ref{SD-section} calculates the isolation probabilities for functions with compact support; Sec \ref{s:methodI} and Sec \ref{s:methodII} provide different methods for calculating isolation probabilities for connection functions with infinite support; Sec \ref{s:numerics} compares approximations with computer simulations and Sec \ref{s:conclusion} concludes the paper. 

\section{Model}\label{s:model}
\subsection{Network Model}
The aim is to understand how boundaries and non-uniformity impact on the global connectivity properties of temporal-spatial networks. 
With this in mind we use a non-uniform Poisson Point Process (PPP) in a triangular region to model the random locations of nodes in the network which represent the locations of people with mobile smart devices.
In particular we focus on a point $\bxi$ located near the corner of the region and study how long it remains isolated from the rest of the network\footnote{$\bxi$ is not in the point process since this would break some of our later assumptions. For example, we will sometimes want to choose $\bxi$ such that it is on the boundary, but often we will also choose the density such that it goes to zero at the boundary}. 
The distribution of nodes is generally taken to be non-uniform which is assumed to be a consequence of human mobility.   
In our calculations we assume the node locations remain fixed throughout the process; this can be interpreted as the system having two different time scales: that  of human mobility, and sending a wireless packet, with the latter being assumed to be much smaller.

Another important assumption is that there is no temporal dependence between time slots, that is to say the probability a node is isolated at time $T$ is independent from the past.

The main metric for discussion will be $\mathbb{P}_{C_T}(\bxi) $, which is the probability a node $\bxi$ has made at least one link to another node in the network in any of the previous time steps. 
For brevity our formulas will be written in terms of the complement of the connection probability $\mathbb{P}_{\text{iso}}^T(\bxi) = 1- \mathbb{P}_{C_T}(\bxi) $, that is the probability $\bxi$ does not make a single connection in any of the previous time slots  $t = 1,2,...T-1, T$.

We now proceed by discussing the point process, distribution of nodes, and the connection functions we adopt in the subsequent subsections. 

\subsection{Point Process}
Let $\Phi$ be a Poisson point process in a region $A$ with non-uniform measure $\Lambda$ with density $\lambda(r, \theta)$, thus the measure $\Lambda(A)$ of a set $A$ is given by $\Lambda(A) = \int_{A} \lambda(r, \theta) r \dd r \dd \theta$. 
The PPP is defined by the following two properties \cite{haenggi2012stochastic}:
\begin{enumerate}
\item For all measurable $A \subset \R^d$, the number of points from $\Phi$ in $A$ ( denoted $\Phi(A)$) is Poisson distributed with mean $\Lambda(A)$,
\item $\Phi(A_i)$ are independent random variables if $A_i$ are mutually disjoint compact subsets of $\R^d$.
\end{enumerate}
Therefore, the probability the number of points in $A$ is $k$ is,  
\es{\mathbb{P}[\Phi(A) = k] = e^{-\int_{A} \lambda(r, \theta) r \dd r \dd \theta}\frac{\left(\int_{A} \lambda(r, \theta) r \dd r \dd \theta\right)^k}{k!}\label{eq:PPP_A}}
In this paper the region $A$ is a right angled triangle determined by $A = \{(x,y): 0\le x \le L, 0 \le y \le x \tan \phi\}$

\subsection{Distribution of points in $\Phi$}
To investigate the impact non-uniformity has we choose a general the density to grow away from the corner,
\es{\lambda(r, \theta) = \bar{N} c r^\alpha g_\phi(\theta), \:\: \alpha \ge 0
\label{eq:f}}
where $\bar{N} $ is the mean number of nodes in the PPP, $c$ is a normalisation constant such that $\int_A \lambda(r, \theta) r \dd r \dd \theta = \bar{N}$ and $g_\phi(\theta)$ can be suitably chosen such that the density  goes to zero on one, both or none of the boundaries. 
One particular example is when  $\alpha = 2$ and  $g_\phi(\theta) = \sin(\theta)\sin(\phi-\theta)$ which approximates the  stationary distribution of the RWP model near a corner.

To approximate the RWP near a corner of a triangle we assume the spatial distribution can be calculated from three independent one-dimensional processes.
The exact expression for the RWP on the line is provided in  \cite{bettstetter2003node} ,\es{f_{1d}(x) = -\frac{6}{L^3}x^2 +\frac{6}{L^2}x,\:\:\:0\le x\le L
\label{eq:rwp1d}}
Thus, making use of the above,  the approximation following the relevant transformations can be written as 
\es{f^{\text{approx}}_{\Delta}(x,y) &= f_{1d}(y) f_{1d}\left(x \cos \left(\frac{\pi}{2} - \phi\right) - y \sin\left(\frac{\pi}{2}-\phi\right)\right)\\
&\times  f_{1d}\left((x-L) \cos \left(\frac{\pi}{2} + \phi\right) - y \sin\left(\frac{\pi}{2}+\phi\right)\right)}
Since we  concern ourselves with what happens near the corner for a large domain, we take the leading order expansion for small $r = \sqrt{x^2 + y^2}$,  
\es{f^{\text{approx}}_{\Delta}(r,\theta) \sim\sin (\theta)\sin(\!\theta\!-\!\phi\!)r^2 + O(r^3)}
So we see that when $g_\phi = \sin (\phi)\sin(\phi-\theta)$ and $\alpha=2$, eq.\eqref{eq:f}  models the RWP model and when $\alpha = 0, g_\phi = 1$ we have the uniform case.

When $\alpha > 0$, even when the domain is taken to be infinite, discussed later,  the expected number of isolated nodes is finite whilst for $\alpha \le 0$ this may not be so \cite{dettmann2017isolation}.
Regardless, we are concerned with the isolation probability of a node near the corner, so approximating the domain to be infinite has little impact and only improves tractability; this is discussed further in the following section.
%Regardless, we are concerned with the local beahviour near teh finite model and the approximations  for $\alpha \ge 0$ we are concerned with thethe overall connectivity will be dominated by those nodes near the corner. 

\subsection{Connection Model} \label{s:Connection_fns}
\begin{figure*}[t]
\centering
%\selectcolormodel{gray}
%\epspdfconversionsetup{gray,bbox=false}
\includegraphics[width = \textwidth, clip = True, trim = {0cm 0cm 0cm 0cm}]{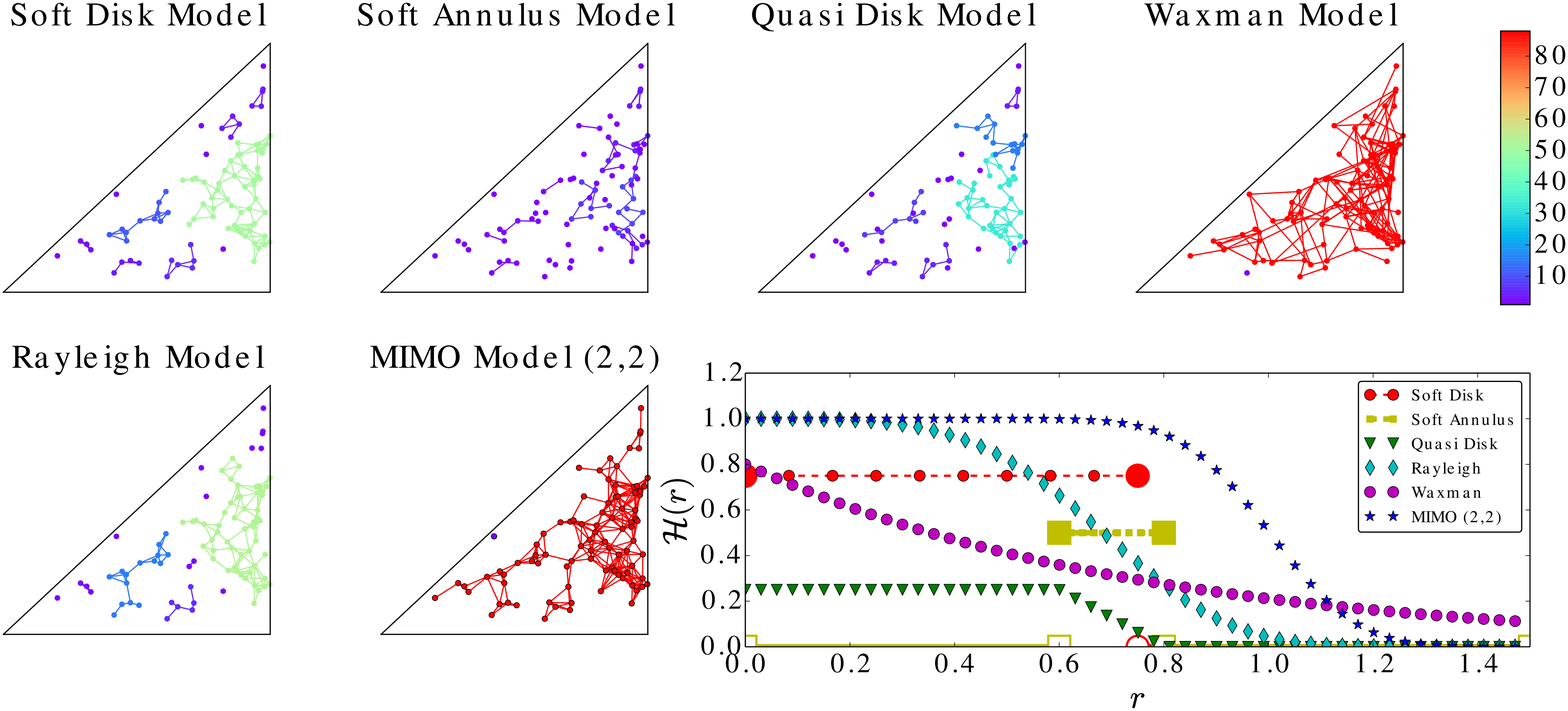}
\caption{A realisation of $\Phi$ for the different connection functions in Table \ref{table:table_H}, and a representation of how the link probability behaves as a function of distance.
In each wedge, the locations of each node are the same, but the links vary.  
The shading represents the size of the connected components of the corresponding graphs with parameters are $\bar{N} = 100; L = 10; r_0 = 0.5; \eta = 4; \beta = 1;  r_-= 0.5; r_+ = 0.8$ and $\wp = 0.75. 0.5, 0.25$ for the SDM, SAM and QDM respectively.
Clearly the Waxman model is the most well connected due to its long range connections, whilst the connectivity in the MIMO model  is better than the Rayleigh case due to multiple antennas. 
}
\end{figure*}

In this paper we consider a range of connection functions controlling the link probabilities, which we assume to have no temporal dependence\footnote{in a more realistic interference model this cannot be assumed\cite{ganti2009spatial}}. 
In general our analysis holds for a wider range of connection functions that are non-increasing, but we focus on those used predominately found in the wireless literature. 

Let $r$ be the Euclidean distance between two nodes in the point process $\Phi$, and thus let $\h(r)$ be the probability these two nodes connect. 
Let $r_0$  be the typical connection range, which is implicit in $\h(r)$ and can be seen in Table\ref{table:table_H}.
For connection functions with compact support, it is typical that only devices that are closer than $r_0$ can form a link, however we also provide variations on this in terms of the Soft Annulus and Qausi Disk models, see below. 
For connection functions with global support  $r_0$ represents how the signal decays resulting in long range connections becoming increasingly unlikely.  
Moreover, $r_0$ can be thought of as a power constraint on transmitting devices and as such we assume the system size to be much larger than the typical connection range, $L \gg r_0$. 
As we consider functions that are non-increasing the significant contributions come from close by which
allows us to take the dimension of the triangle to be infinite, referred to as a wedge ($\mathcal{W}$), exploited in Sec \ref{s:methodI} and \ref{s:methodII}, without losing much accuracy.
%{\CB
To investigate the impact of boundaries we assume the node $\bxi$ is located near the corner, and that $|\bxi| < r_0$.
%} 
This is to simplify calculations involving connection functions with finite support.  More general calculations are straightforward but cumbersome, and do not provide greater insight.

We define the following seven connection functions in Table \ref{table:table1}, and discuss the connection functions with compact and infinite support separately below,see Ref \cite{dettmann2016random} for more background.

\begin{table}
\centering\caption{}
    \begin{tabular}{ | l | l |}
    \hline
    Model & $\h(r)$ \\ \hline
    SDM & $\wp \mathbbm{1}_{r \le r_0}$ \\ \hline
    SAM & $\wp\mathbbm{1}_{r_- \le r \le r_+} $  \\ \hline
    %PLM & $\wp \left(1-\left(\frac{r}{r_0}\right)^\mu\right)\mathbbm{1}_{r \le r_0} $ \\     \hline
    QDM & $\begin{cases}
\wp\:\: & 0 \le r \le r_- \\
\wp - \wp\left(\frac{r - r_+}{r_+ - r_-}\right)^\mu\:\: & r_- \le r \le r_+\\
0\:\: &\text{otherwise}
\end{cases}$  \\
    \hline
    Rayleigh & $e^{-\left(\frac{r}{r_0}\right)^\eta}$  \\ \hline
    Waxman & $\beta e^{-\frac{r}{r_0}}$  \\ \hline
    Interference & $e^{-q\sigma^2 r^{\eta}}e^{-\int_{\mathcal{W}} \frac{q \zeta r^{\eta}}{|\boldsymbol{z}|^\eta + q\zeta r^{\eta}}\Lambda( \dd \boldsymbol{z})}$ \\
    \hline
    MIMO & $e^{\!-\left(\!\frac{r}{\!r_0}\!\right)^{\!\eta}}\left(\!2\!+\!\left(\!\frac{\!r}{\!r_0}\!\right)^{\!2\eta}\!-\!e^{-\!\left(\!\frac{r}{r_0}\!\right)^{\!\eta}}\right)$ \\
    \hline
    \end{tabular}
    
    \label{table:table_H}
    \vspace{0.5cm}
    Table I: Table of connection functions which are discussed in \ref{s:Connection_fns}. Parameters: $r_0$ is the typical connection range, $\wp \in (0,1]$ is the probability a node is active; $\eta \in [2, 6]$ is the path loss exponent; $\mu >0$ defines how fast the function decays with distance; $q>0$ is the threshold signal quality and the noise in the channel is given by $\sigma^2$; 
  \end{table}

\subsubsection{Connection Functions with finite support} 
The soft-disk model (SDM) is a variation on the random geometric graph (RGG), introduced in \cite{gilbert1961random}.
Two nodes form a link with probability $\wp \in (0,1]$ if their Euclidean separation $r\le r_0$.
The probability $\wp$ is used throughout this paper to incorporate a temporal aspect into the models with compact support; with $\wp = 1$ we have a deterministic model and no temporal aspect, the  case $\wp = 0$ is excluded as every node is isolated.
The nodes in $\Phi$ (equivalently links) can be thought of as becoming active with probability $\wp$.
 
The Soft-annulus(SA) model  is a modified version on the SD model where links can only be formed in the interval $r  \in [r_-, r_+]$.
Intuitively  this exclusion region  can be seen as a simple channel access scheme ensuring two nodes in close proximity transmit on different channels (thus cannot connect to each other) in order to mitigate interference effects.

The Quasi Disk  (QD) model is the first model we discuss that models the connection probability decaying with distance. 
The QD model is a piecewise connection model that has support on $r \in [0, r_+]$ and combines the SD model and one which decays with $r$; for $r \in [0, r_-]$ the connection probability is $\wp$, whilst for  $r \in (r_-, r_+]$ the connection probability decays to zero. 
The factor $\wp$ is included throughout to ensure $\h(r)$ is continuous at $r_-$, whilst the parameter $\mu$ is used to tune how "fast" the connection probability decays to zero, with it doing so faster for small values of $\mu$. 
Notice that by taking $r_- = 0$ the connection probability decays to zero with $r$, whilst it reduces to the SD model when $r_- = r_+$.
Intuitively the reader can think of  QD model to represent a connection environment which is clutter free within the ball $B(\bxi,r_-)$, whilst the signal decays between $r_-$ and $r_+$ due to the appearance of obstacles.
%the fading increases due to more obstacles. 
Alternatively, the inner ball could model a region where all transmissions are done on separate channels whilst channels are shared in the outer ball creating interference.

\subsubsection{Connection Functions with infinite support} 
One of the  most widely used connection functions in wireless communications (which has a similar analgoue in neral networks  \cite{shiino1992self}) is the probability that the Signal-to-Noise-Ratio (SNR) is greater than some threshold $q$. 
By modelling the signal as the product of channel gain $|h|^2$ (an exponential random variable with mean $1$ which models the small scale fluctuations in the channel) and  pathloss function $r^{-\eta}$ (which models how the signal decays with distance), and taking the noise to be $\sigma^2$ one can show that the connection function behaves like a stretched exponential, with a scaling $r_0^{-\eta} = \frac{1}{q \sigma^2}$.
More specifically, 
\est{\h = \mathbb{P}[\text{SNR} > q] = \mathbb{P}\left[\frac{|h|^2 r^{-\eta}}{\sigma^2} > q\right] =\exp\left(- \frac{q\sigma^2}{r^{-\eta}}\right)
} 
Empirical observations have shown that typically $\eta \in [2,6]$ in urban environments \cite{dettmann2016random}, when $\eta = 2$ the signal decays like that in free-space, where as in cities there is less likely to be long range connections due to obstacles thus $\eta$ will generally be larger.
In highly reflective mediums $\eta < 2$. 

The Waxman case, see Table \ref{table:table_H}, is closely related to the Rayleigh  model, where connections are very soft. 
The Rayleigh model reduces to the Waxman model for $\eta = 1$, and assuming $\beta = 1$. 

The MIMO (Multiple input and multiple output) connection function models the case when  the receiver and transmitter have multiple antennas. 
Due to the limited battery power of mobile devices, the number of antennas is unlikely to be large so we focus on the case when each device has two input and two output antennas, and the channels are assumed to be i.i.d Rayleigh channels.
Work on a more general array of antennas can be found here \cite{dettmann2016random,kang2003largest}.

\begin{figure*}[t]
\includegraphics[scale = 0.35]{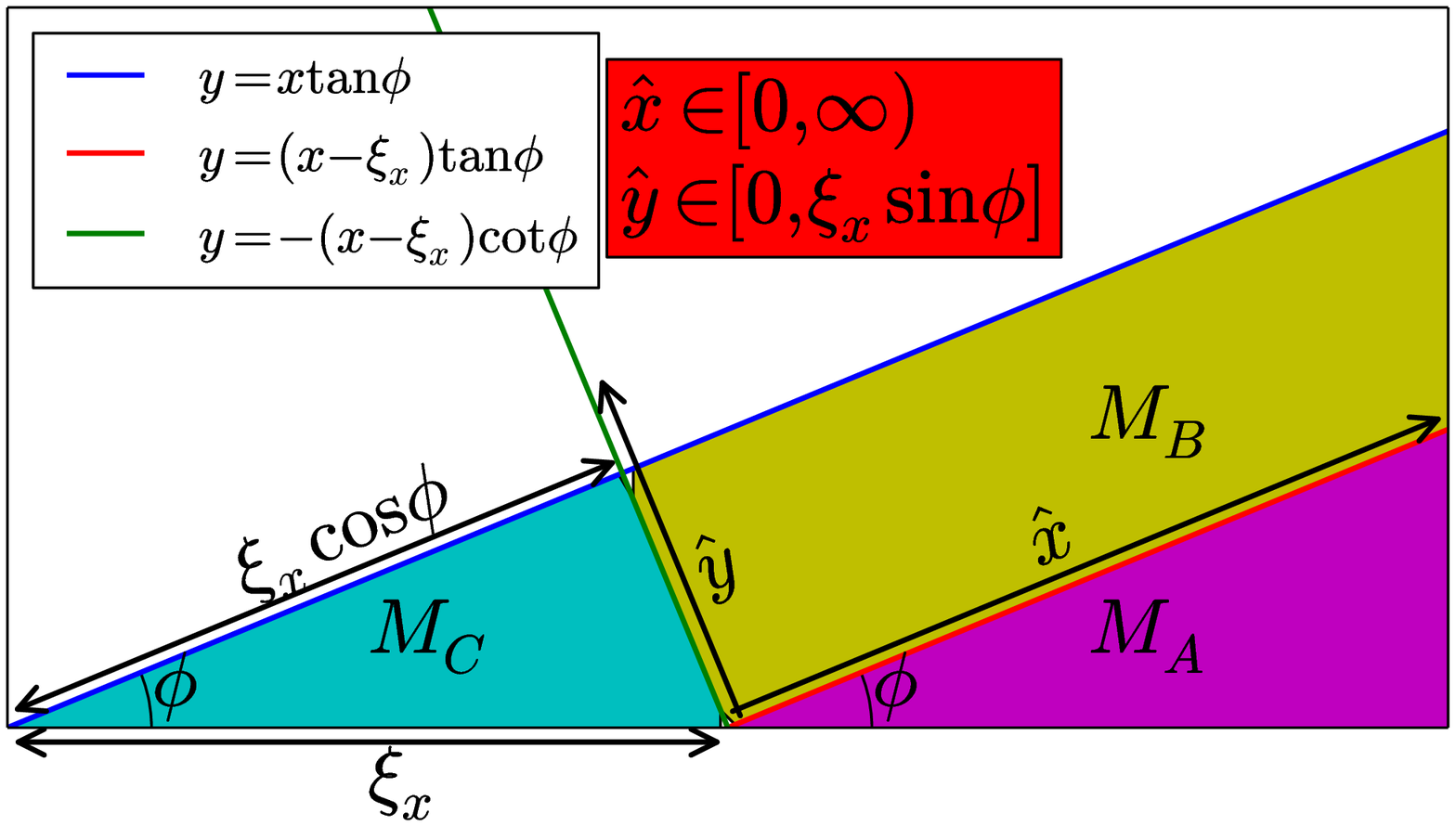}\includegraphics[scale = 0.35]{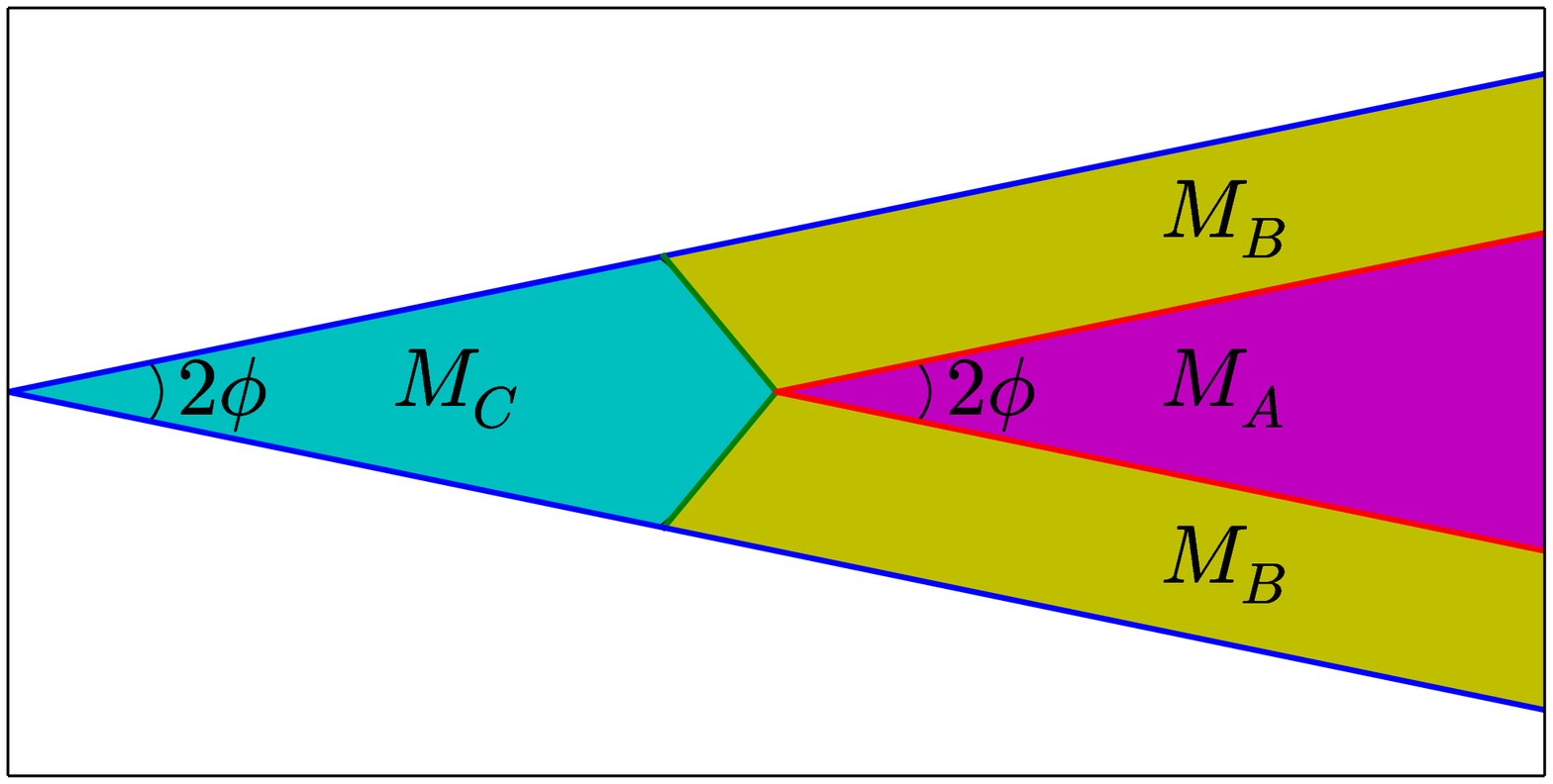}
\caption{(Left): Schematic of the wedge and the regions $M_a, M_b$ and $M_c$. (Right) By combining two wedges together we can calculate the probability that a single user is isolated from the network near the corner. }
\label{fig:schematic}
\end{figure*}

Finally the last connection function we consider is one that includes interference, where the noise $\sigma^2$ is negligible.
%{\CR example of scenario}
Due to links being dependent on the number and  locations of other nodes in $\Phi$,  the network can become highly directional unlike the other models previously discussed; the probability a node $\bxi$ can successfully transmit a message to $\by$ is distinct from the probability it can receive a message from $\by$.
For simplicity we consider the latter as the interference is measured at the  receiver $\bxi$. 

We proceed by giving the general definition for the connection probability between a transmitter in $\Phi$, $\bX_{\mathcal{T}} = (X_{\mathcal{T}}, \theta_{\mathcal{T}})$, and receiver $\bX_{\mathcal{R}} = (X_{\mathcal{R}}, \theta_{\mathcal{R}})$, (the receiver is assumed not to be in $\Phi$), with $r = |\bX_{\mathcal{T}} - \bX_{\mathcal{R}}|$ being the point to point distance of the link.
Denote the interfering nodes in $\Phi$ as $\bX_{\mathcal{I}} = (X_{\mathcal{I}}, \omega_{\mathcal{I}})$. 
\es{\h(r)\!&=\!\mathbb{P}[\text{SINR}\! >\! q]\!\\
&=\mathbb{E}\left[\!\mathbb{P}\!\left[\frac{|h_{\tau}|^2 r^{-\eta}}{\zeta\!\! \sum\limits_{\bX_{\mathcal{I}} \in \Phi\backslash{\bX_{\mathcal{T}}}}\!\!\!\!\!|h_k|^2 |\bX_{\mathcal{I}}\! -\! \bX_{\mathcal{R}}|^{-\eta}} > q\biggl|\Phi\right]\right]\\
&=\!\! \exp\left(\!-\!\bar{N}\int_{\mathcal{W}} \!\!\frac{\lambda(z, \omega) z}{1\!+\! \frac{(z^2 +X_{\mathcal{R}}^2 - 2 z X_{\mathcal{R}} \cos(\theta_{\mathcal{R}} - \omega_{\mathcal{I}}))^{\frac{\eta}{2}}}{q \zeta r^\eta}}\!  \dd z \dd \omega \right)
\label{eq:laplace}
} 
In the third equality we have used that the channel gain $|h_k|^2$ is an i.i.d exponential random variable, and used the PGFL to average over all possible locations of the interferers \cite{haenggi2012stochastic}. 
It is often the case that eq.\eqref{eq:laplace} cannot be given in closed form for an arbitrary location of $\bxi$ in finite domains with non-uniform measure.% due to a loss in stationarity.
In section \ref{Int_section} we will make several approximations to allow for a more tractable analysis. 
 
The interference model is the only connection model that depends on the underlying point process. The other connection models can be thought of as having networking protocols which mitigate the impact of interference, hence the restriction that $r_0 \ll L$ due to a finite amount of network resources.

\section{Isolation Probabilities}\label{s:iso_probs}
In this section we provide three methods for computing the probability a node $\bxi$ is isolated for $T$ consecutive time slots near a corner. 
The first method is applied to connection functions with compact support, whilst the other two are used for connection functions with global support. 
The last two methods can also be applied to those connection functions with compact support and the corresponding discontinuities can be handled separately although these contributions can often be ignored in the small parameter expansions \cite{dettmann2017isolation}. 
We proceed by giving the initial formulation of the analysis, and then consider each method separately in the subsequent subsections. 

The probability that a user $\bxi$ is isolated from all other points in $\Phi$, conditioned on $\Phi$, for $T$ consecutive time steps is,
\es{
\mathbb{P}_{\text{iso}}^T(\bxi|\Phi) &=  \prod_{\by \in \Phi} \left(1 - \h\left(|\bxi - \by|\right)\right)^T,
\label{eq:PT}}
By averaging over all possible realisations of $\Phi$, and using the probability generating functional for poisson point processes \cite{haenggi2012stochastic},
\est{G(v) = \mathbb{E}\left[\prod_{\zeta\in\Phi}v(\zeta)\right] = \exp\left(- \int (1-u(\zeta))\Lambda(\dd \zeta)\right),}
we can write eq\eqref{eq:PT} as,
\es{\mathbb{P}_{\text{iso}}^{T}(\bxi)=  \exp\left(-\int_{A}\left(1 - \left(1 - \h\left(|\bxi - \by|\right)\right)^T\right)\Lambda(\dd \by)\right)
\label{eq:iso}}
where the integral is over the triangular region, and $\Lambda$ is the intensity measure of $\Phi$.
For a single time slot eq \eqref{eq:iso} reduces to, 
\es{\mathbb{P}_{\text{iso}}^{T=1}(\bxi)=  \exp\left(-\int_{A} \h\left(|\bxi - \by|\right) \lambda(\by)\dd \by\right) = e^{-M(\bxi)}
\label{eq:iso_t_0}}
where $M(\bxi)$ is the usual connectivity mass, \cite{coon2012full, dettmann2016random}

As an aside, it turns out that if mobility is included between time slots, the average time it takes for $\bxi$ to connect decreases \cite{haenggi2013local ,grossglauser2001mobility}.
For example, as a crude lower bound we can consider the case where each node in $\Phi$ has infinite mobility,  i.e there is no spatial correlation in the location of nodes from one time step to another,then the probability a node is isolated for $T$ consecutive time steps is simply %{\CR $M(\bxi)^T$}
$e^{-T M(\bxi)}$.
The probability a node $\bxi$ is connected at time $T$ is the CDF of a geometric random variable with mean  $M(\bxi)$, therefore $\bxi$ can always transmit in finite time provided $M(\bxi)>0$.
In this model the number of points in $\Phi$ during each time slot is a random variable with mean $\bar{N}$ and can be thought of as nodes randomly turning on/off. 
Alternatively, one could condition on the number of points in each time step by using the Binomial Point Process.

For a fixed stationary distribution of nodes we return to  eq.\eqref{eq:iso}.
Consider the limit as $T\to \infty$ for any $\h(r) > 0$ with infinite support, in a finite domain $A$,
\es{\mathbb{P}_{\text{iso}}^T(\bxi) =^{T\to\infty} \exp\left(- \int_{A} \lambda(r,\theta) r \dd r \dd \theta\right) = e^{-\bar{N}} > 0}
The probability that a node $\bxi$ is isolated is always positive, since the probability the point process $\Phi$ is empty ($e^{-\bar{N}}$) is also positive, a finite domain effect.
A similar analysis holds for connection functions with compact support, but instead  $\mathbb{P}_{\text{iso}}^T(\bxi)$ equals the probability the region where links can be made is empty. 
Consequently, the local mean in/out delay (average time it takes for a node to transmit a packet) is infinite for finite networks. 
This can also be the case for infinite networks where the connection function is $\h(r) = e^{-\left(\frac{r}{r_0}\right)^\eta}$ which is a result of the appearance of arbitrarily large voids in the network  \cite{baccelli2011optimal}.
%This holds for non-uniform networks. see notes. 
This behaviour can be mitigated in both cases by conditioning on a point being a distance $d<\infty$ away, or in the case of a finite network and unbounded support fixing the number of points. 

We proceed by using eq.\eqref{eq:iso} to calculate the isolation probabilities for different connection functions expressed in Sec \ref{s:Connection_fns}, starting with those with compact support. 

\subsection{Connection functions with compact support}
The method used for calculating the isolation probability (and thus connection probability) for $\bxi$ is very similar for all models (with the exception of the Quasi-disk case which is discussed in Sec\ref{s:methodII}) so we proceed by deriving it for the Soft Disk model, and give the results for the SA model in Table \ref{table:table1}.

\textit{Example: Soft Disk Model}\label{SD-section}

From eq.\eqref{eq:iso} it is not possible to obtain an explicit expression not in terms of integrals  for $\mathbb{P}_{\text{iso}}^T$ when $\bxi$ is located at an arbitrary location in $A$. 
However, in this paper we concern our analysis with the particular case when $\bxi = (x, \omega)$ isolated near the corner, and $r_0 \ge x$ which guarantees that the ball centred at $\bxi$ with radius $r_0$, $B_{\bxi}(r_0)$, intersects both boundaries and includes the vertex at the origin.   
From these assumptions, and eq.\eqref{eq:iso} we have, 
\es{\mathbb{P}_{\text{iso}}^T((x, \omega)) &= \exp\left(- \left(\!1\!-\!\left(\!1\!-\!\wp\!\right)^T\!\right)\!\int_0^\phi\!\int_0^{z} \lambda(r, \theta) r \dd r \dd \theta\right)\\
\mathbb{P}_{\text{iso}}^T((x, \omega)) &= \left(\exp\left(- \!\int_0^\phi\!\int_0^{z} \lambda(r, \theta) r \dd r \dd \theta\right)\right)^{\left(\!1\!-\!\left(\!1\!-\!\wp\!\right)^T\!\right)}\\
\mathbb{P}_{\text{iso}}^T((x, \omega)) &= V_{\mathcal{B}}(\bxi, r_0)^{\left(\!1\!-\!\left(\!1\!-\!\wp\!\right)^T\!\right)}\label{eq:SDM}}
where $z = \sqrt{r_0^2 + x^2 - 2 r_0 x \cos(\theta - \omega)}$, and $V_\mathcal{B}(\bxi, r_0)$ is the void probability , the probability there is no node in the ball ($\mathcal{B}(\bxi, r_0)$) of radius $r_0$ centred at $\bxi$ in $A$ which is directly computed by setting $k= 0$ in eq.\eqref{eq:PPP_A}.  
For the uniform case ($\alpha = 0, g_\phi(\theta) = 1$) the inner integral in eq.\eqref{eq:SDM} is proportional to the size of the region.
For the general case we expand the integrand of eq\eqref{eq:iso} for small $x (\le r_0)$ to provide a closed form approximation,
\es{
\mathbb{P}_{\text{iso}}^T((x,\omega)) &= \exp\biggl(-\bar{N} c(1-(1-\wp)^T)\biggl(\frac{r_0^{\alpha + 2}}{\alpha + 2} G_\phi \\&- G_c(\omega) r_0^{\alpha + 1} x + G_2(\omega) r_0^{\alpha} x^2 \biggl) \biggl)
\label{eq:SDM_approx}} 
where $G_\phi = \int_0^\phi g_\phi(\theta) \dd \theta $, $G_c(\omega) = \int_0^\phi g_\phi(\theta) \cos(\theta) \dd \theta$, $G_2(\omega) = \int_0^\phi \frac{1}{2}g_\phi (1 + \alpha \cos^2(\theta - \omega)) \dd \theta $.
At the corner the above reduces to just taking the leading order term.
See Table \ref{table:table1} for a similar expression for the soft-annulus model.

In the limit as $T\to \infty$ we return to the original void probability, for the SA model it converges to the probability the annulus $V_{\mathcal{A}}(\bxi,r_-, r_+)$ is empty.
We notice that this type of connection function with compact support results in no guarantee that $\bxi$ connects, even if the PP is non-empty as the relevant connection region might be; trivially this \textit{all or nothing} type of connection means we need at least the average number of nearest neighbours to be greater than one\cite{gilbert1961random}.

\subsection{User Isolation - Method I}\label{s:methodI}
In this section we focus on connection functions with global support, and provide a method based on translating the distance between points, since local behaviour will dominate (very long connections are unlikely)  we approximate the domain to be infinite for tractability. 

We first start by writing eq.\eqref{eq:iso} as 
\es{\mathbb{P}_{\text{iso}}^T(\bxi) &= \!\exp\left(-\!\!\int_0^\phi\!\int_0^{\frac{L}{\cos \theta}}\bar{\h}\left(z\right)\lambda(y,\theta)y  \dd y \dd \theta\right)\label{eq:gen},}
where the node $\bxi$ is located (in polar coordinates) at $(x, \omega)$,$\bar{\h}^T\left(z\right) = \left(1\! -\!\left(1\!-\! \h\left(z\right)\right)^T\right)$  and  $z = \sqrt{x^2+y^2 - 2 xy \cos(\theta-\omega)}$ is the corresponding transformation using the cosine rule. 
By assuming discrete time we can expand the integrand using the binomial theorem,  expand for small radial component x and assume the contributions come from near by so the domain is assumed to be infinite to give, 
\es{\mathbb{P}_{\text{iso}}^T(\bxi) &=  \exp\biggl(-c \bar{N} \sum_{k=1}^T (-1)^{k+1} {T \choose k}  \biggl(\h_{k,\alpha+1}G_\phi\\
&- k \frac{1}{r_0} \h^\prime_{k-1,\alpha+1}x G_c(\omega) + O(x^2) \biggl) \biggl)
\label{eq:gen2}}
where  $G_\phi, G_c(\omega)$ are as before, 
$\h^{(n)}_{k,\alpha} = \int_0^\infty \h^{(n)}\left(\frac{y}{r_0}\right) \h^k\left(\frac{y}{r_0}\right) y^{\alpha} \dd y $ and $^{(n)}$ corresponds to $n^{th}$ derivative. 

We now proceed by calculating the isolation probabilities for the Rayleigh and Interference connection functions outlined in section \ref{s:Connection_fns} through direct application of eq.\eqref{eq:gen2}. \\

\textit{Example I: Rayleigh Connection Model}

First we consider the Rayleigh connection function defined in Table \ref{table:table_H} and through eq.\eqref{eq:gen2} we obtain,
\es{
-\frac{\log \mathbb{P}_{\text{iso}}^T(\bxi)}{c \bar{N}} &= \frac{r_0^{\alpha+2}}{\eta}\Gamma\left[\frac{2+\alpha}{\eta}\right]H_{T,1}^{\frac{2+\alpha}{\eta}}G_\phi\\
&+  r_0^{\alpha+1} \Gamma\left[\frac{1+\alpha}{\eta}\!\! +\! 1\right]H_{T,1}^{\frac{1+\alpha}{\eta}}x G_c(\omega) +...  \\
\label{eq:rayleigh}}
where $H_{T,\beta}^{s} = \sum_{k=1}^T (-1)^{k+1} {T \choose k} k^{-s} \beta^k	 $ is the generalised Roman harmonic number given in \cite{roman1992logarithmic,dettmann2017isolation}.
Note that we include the constant $\beta$ for the Waxman case, the result of which is given in Table \ref{table:table1}.
Using an asymptotic approximation provided in \cite{dettmann2017isolation}, we can approximate the isolation probabilities for large $T$, where $\gamma$ is the Euler-Mascheroni constant, 
\es{H_T^s\!\approx\!\frac{(\log T)^s}{s}\!+\!\gamma(\log T)^{s-1}\!\! +\!\! \frac{(6 \gamma^2\! +\! \pi^2)(\!s\!-\!1\!)}{12}(\log T)^{s-2} + ...
\label{eq:harm_asymptotic}}
This provides a good match when $s \le$  which implies for more cluttered environments (higher value of $\eta$) the approximation improves; for $s = 1$ we obtain the standard harmonic number.
The above approximation can be rescaled to include a constant $\beta$ by replacing $\log T$ with $\log(\beta T)$.

Conversely, when $\frac{\alpha + 2}{\eta}$ is large, we have,
\es{\int_0^\infty\!(1\!-\!(1-\!e^{-z})^T)\!z^{\frac{\alpha+2}{\eta} - 1}\!\dd z &\sim T \Gamma\left[\frac{\alpha+2}{\eta}\right]
\label{eq:harm_asymptotic2}}
This is a useful approximation for a very inhomogeneous network (or a highly reflective environment) and suggests the isolation of nodes after time $T$ slots behaves like $\exp(- \text{constant} T)$. 
When $\eta =1$ (Waxman model) the above provides a good approximation, particularly for the RWP distribution. 
\textit{Remark:} The exact transition behaviour between the two regimes is more subtle and not studied here.

\textit{Interference}\label{Int_section}

In Sec.\ref{s:Connection_fns} we introduce the interference limited connection function for a node with a general location in $A$. 
A tractable form of $\h$ (not expressed in terms of hyper geometric functions) is only possible for the specific case when $\bxi$ is located at the corner and the domain is assumed to be infinite (Wedge). 
\es{\h(r) &= \exp\left(-\int_0^\phi\int_0^{\frac{L}{\cos\theta}} \left(1 - \frac{1}{1 + \frac{q \zeta r^\eta}{z^\eta}}\right)\lambda(z,\theta)z\dd z \dd \theta\right)\\
&=^{(*)}\!\frac{ c \bar{N}s^{\frac{2+\alpha}{\eta}}\pi}{\eta}\csc\left(\!\frac{(2+\alpha)\pi}{\eta}\!\right)G_\phi\\
&= c \bar{N} G_\phi c_{\mathcal{I}} r^{2+\alpha}}
where $*$ denotes we have assumed an infinite wedge, $c_{\mathcal{I}} = \frac{ (q \zeta)^{\frac{2+\alpha}{\eta}}\pi}{\eta}\csc\left(\frac{(2+\alpha)\pi}{\eta}\right)$  and we require $\alpha + 2 < \eta$ to hold. 
The condition that $\alpha + 2 < \eta$  ensures that there is indeed positive probability of connection; there exists a phase transition at $\eta =$ dimension such that for $\eta \le $ dim the global behaviour begins to  dominate and the aggregate interference causes disconnection.
Since we assume an infinite wedge, which has an infinite number of nodes, we need to ensure the  \textit{local behaviour} dominates, hence $\alpha + 2 < \eta$.
Consequently, for the RWP case we need $\eta > 4$; i.e a "very" urban environment like a large city such as New York.
Alternatively we can make the approximation that all non-negligible interference comes form all those devices within a distance $r_{\mathcal{I}}$ which allows for the relaxation of the $\alpha+2 < \eta$ restriction;  but yields a connection function in terms of hyper geometric functions which leads to an intractable calculation later; see \cite{di2016intensity} amongst others on approximating interference.

When the node is located near the corner we can compute the approximation through method I or II (outlined in the next section).
For method I we approximate the connection probability at $x$ to be the same as at the vertex a the origin such that we can apply eq\eqref{eq:gen2}, noting $r_0 = 1$, to get 
\es{
-\log \mathbb{P}_{\text{iso}}^T(\bxi) &=\!\frac{H_T}{(2+\alpha)c_{\mathcal{I}}}\\
&+\!\frac{(c \bar{N})^{\frac{1}{2+\alpha}}}{(c_{\mathcal{I}} G_\phi)^{\frac{1+\alpha}{2+\alpha}}}\Gamma\left[\!\frac{3+2\alpha}{2+\alpha}\!\right]H_T^{\frac{1+\alpha}{2+\alpha}} x G_c(\omega) + O(x^2) 
\label{eq:InterferenceAP}}
where $H_T^s$ is the Roman harmonic number defined earlier and $H_T$ is the usual harmonic number with asymptotic expansion
\es{ H_T
&=  \log T + \gamma + \frac{1}{2T} - \frac{1}{12 T^2} + O(T^{-4})
\label{eq:iso_int_no_noise_3}} 
The leading order term in  eq.\eqref{eq:InterferenceAP} is independent of the density of users and the angle of the wedge. 
This is consistent with the results in \cite{andrews2011tractable} which highlights how any increase in signal power due to proximity is counter balanced by an increase in the interference field.

However, the second term (first order correction term) scales like $\bar{N}^{\frac{1}{2+\alpha}}$ and does in fact depend on both the geometry of the wedge and the density of users, ultimately leading to  $\lim_{\bar{N}\to\infty} \mathbb{P}_{\text{iso}}^T(\bxi) \to \mathbb{P}_{\text{iso}}^T(\underline{0})$.
Intuitively this is because in the high density limit\footnote{this is only true for our particular choice of path loss model \ref{s:Connection_fns} \cite{pratt2016does}},  the local picture for each node looks the same due to the scaling of power and interference which means connections are dominated by local nodes ( assuming $\alpha +2 < \eta$). 
\subsection{User Isolation - Method II}\label{s:methodII}
In this section our aim is to give an alternative approach to Method I which provides greater tractability and  is more suited to more complicated connection functions $\h(r)$.
As such,this method, Method II, is more suited to more complicated connection functions such as MIMO or those outlined in \cite{dettmann2016random} where closed form expressions cannot be obtained via method I, or else the computation of the higher order moments of the connection function are time consuming.   
For a non-increasing connection function $\h(r)$ with global support the approximation can be expressed as a combination of one-dimensional integrals which are quick to numerically compute.
In this analysis we require the  density to go to zero along the top border, which is akin to the RWP case or other mobility models where boundaries are left largely unexplored. 

In this section we will consider the user located on the bottom boundary, $\bxi = (\xi_x, 0)$, and divide the domain into three regions $M_A, M_B$ and $M_C$, see Fig.\ref{fig:schematic}, such that, 
\est{\mathbb{P}_{\text{iso}}^T(\bxi = (\xi_x,0))&= \exp\left(-(M_A + M_b + M_c)\right)} 

To obtain an expression for a user located near the corner, but not on either boundary, we can combine two triangular domains together along the non-zero boundaries to obtain  $\mathbb{P}^T_{\text{iso}}[\xi_x,\omega]$. 
In general the two triangular regions not be identical, but we assume so merely for brevity.
We now proceed to calculate each of the contributions from these sub-regions using eq.\eqref{eq:iso}, starting with $M_A$ . 
\subsubsection{Region $M_A$}
The region $M_A$, as shown by the purple region in Fig\ref{fig:schematic}, has a transformed polar coordinate system centred at $(\xi_x, 0)$. 
For this case we use the cosine rule to make the necessary transformation of the density.
\es{M_A &=\int_0^{\phi} \int_0^{\frac{(L- \xi_x)}{\cos \theta}} \bar{\h}^T \lambda(z, \hat{\omega}) y \dd y \dd \hat{\theta}\\
\!&=\!\!\int_0^{\infty}\!\bar{\h}^T\left(y^{\alpha+1}\!G_\phi\!+\!\left(\!\alpha\!+ \!1\right) y^\alpha G_c(0) \xi_x + ... \right) \dd y
\label{eq:M_A}}
where $z = \sqrt{y^2 + \xi_x^2 - 2 y  \xi_x \cos(\pi- \hat{\theta})}$, $\hat{\omega} = \arcsin \left[\frac{y \sin \hat{\theta}}{\sqrt{y^2 + \xi_x^2 + 2 y \xi_x \cos \theta}}\right]$ and  $\bar{\h}^T = \!1-\!\left(\!1-\!\h\!\left(y\right)\!\right)^{\!T}\!$. 
In the above, we have expanded for small $\xi_x$ and  assumed an infinite wedge. 

In fact the main contribution arises form the region $M_A$ as we will see in the following subsections as  the contributions form other regions are of order $\xi_x^2$.
\subsubsection{$M_B$}
The region $M_B$ is coloured yellow in Fig\ref{fig:schematic} and has a translated and rotated coordinate system $(\hat{x}, \hat{y})$. 
Throughout this section, since the function $g_\phi(\theta)$ goes to zero near the border we approximate $\hat{y}$ as small. 
The connection function can therefore be approximated as, 
\es{\!\h\left(\!\!\sqrt{\hat{x}^2 + \hat{y}^2}\right)^k \! &\! \approx \h^k\left(\!\hat{x}\right)\!+\!\frac{k}{ \hat{x}}\h\left(\!\hat{x}\right)^{k-1}\!\!\h^\prime\left(\!\hat{x}\right)\hat{y}^2\!+\!...\\
}
Using this approximation, and assuming the discrete time so we can rewrite the integrand as a sum, we have that the contribution from the region $M_B$ is, 
\es{M_B &= \int_{M_B}\bar{\h}\left(|\bxi - \by|\right)^T \lambda(y,\theta) y\dd y \dd \theta\\
\!&\approx\!\frac{\bar{N}cg_\phi^\prime(\phi)}{2} \sum_{k=1}^T {T \choose k}\!(-1)^{k+1}\!\xi_x^2\!\sin^2\phi \h_{k,\alpha-1}\left(\hat{x}\right)+ o(\xi_x^2) \\
}
We notice immediately that the leading order term is indeed of order $\xi_x^2$ which we will neglect from our final approximation.  

\subsubsection{$M_C$}
For the $M_C$ region, neighbouring nodes are close by so we approximate $\h(r) \approx 1$, and we observe that the contribution is proportional to the size of the region, 
\es{M_C &= \int_0^\phi\!\int_0^{\frac{\xi_x \cos \phi}{\cos(\phi+\theta)}} \left(1-\left(1-\h\left(r\right)\right)^T\right)\lambda(r,\theta)r\!\dd r\!\dd \theta\\
&= \int_0^\phi\!\int_0^{\frac{\xi_x \cos \phi}{\cos(\phi+\theta)}} \bar{\h}^T\left(r\right) \lambda(r,\theta)r\!\dd r\!\dd \theta\\
&\approx  \bar{N} c\frac{\xi_x^{2+\alpha}}{2+\alpha} \int_0^\phi  g_\phi(\theta) (\cos \phi \sec (\theta-\phi))^{2+\alpha} \dd \theta}
In fact the best case scenario (in this particular model) is for the uniform distribution,where  $\alpha = 0$, and $g_\phi(\theta) = 1$ leaving, 
\est{M_C = \bar{N} c \frac{\sin (2 \phi)}{4}\xi_x^2}
By combining the contributions from each region and taking terms up to order $\xi_x$ the probability a point located along the border is isolated can be written in terms of the following simplified 1-dimensional integral,
\es{-\frac{\log \mathbb{P}^T_{\text{iso}}[(\xi_x,0)]}{c \bar{N}}\!&\!\approx\!\int_0^\infty \! \bar{\h}^T\! \left(\!r^{\!\alpha\!+\!1} G_\phi +\!(\alpha \!+ 1)\! G_c r^\alpha\!\xi_x\!+\!...\right)\!\dd r , 
\label{eq:Method2}}
We now have the integral in the form, with a change of variables, 
\est{I(s) = r_0^{s+1}\int_0^{\infty} \left(1- \left(1- \h\left(r\right)\right)^T\right)r^s \dd r}
and for the asymptotic approximations we need only expand once for large $T$ and we are done.
This method provides a greater tractability since it involves computing only one integral (albeit with different parameters $s$), and for large times often an asymptotic approximation can be found.

We now proceed by computing the isolation probabilities for the MIMO and Quasi-disk connection functions.

\textit{Example: MIMO}

For the MIMO connection function we apply eq. \eqref{eq:Method2} directly.
\es{-\frac{\log \mathbb{P}_\text{iso}^T(\bxi)}{\bar{N}c} &=
\frac{r_0^{\alpha+2}G_\phi}{\eta} \int_0^{\infty} \bar{\h}^T r^{\alpha+1} \dd r\\
&+ \frac{(\alpha +1)r_0^{\alpha+1} G_c  \xi_x}{\eta} \int_0^{\infty} \bar{\h}^T r^\alpha \dd r \\
&=\frac{r_0^{\alpha+2}G_\phi}{\eta} I_1\left(\frac{\alpha+2}{\eta}\right)\\
&+ \frac{(\alpha +1)r_0^{\alpha+1} G_c \xi_x}{\eta} I_1\left(\frac{\alpha+1}{\eta}\right)
\label{eq:M_A_MIMO}}
where the integral is $I_1(s) = \int_0^\infty (1 - (1- e^{-x}(2 + x^2 - e^{-x})^T)x^{s-1}\dd x$.
First we consider the case when $\alpha$ is large (equivalently s large); and $T$ small in comparison, we can get a simple expression for the asymptotic behaviour. 
\es{I_1(s)&=\int_0^\infty \left(1 - \left( 1- e^{-x}(2 + x^2 - e^{-x})\right)^T\right)x^{s - 1} \dd x\\
&\sim 2 T \Gamma\left[s\right] + T \Gamma\left[s+2\right] - T2^{-s}\Gamma\left[s\right]} 
For  $s \le 1$, which will often be the case, we can do a similar asymptotic expansion to \cite{dettmann2017isolation} by splitting the integral up at $\hat{c}\log T$, where $\hat{c}$ is a constant. 
\es{I_1(s) &= \int_0^\infty \left(1 - \left( 1- e^{-x}(2 + x^2 - e^{-x})\right)^T\right)x^{s-1}\dd x\\
 &= \frac{\hat{c} \log^s T}{s} + (\log [T^{1-\hat{c}}(\hat{c} \log[T])^2] + \gamma) (\hat{c} \log[T])^{s-1}\\
&+ \biggl(6\gamma^2 + \pi^2 + 12\gamma \log[T^{1-\hat{c}}(\hat{c} \log[T])^2]\\
&+ 6(\log[T^{1-\hat{c}}(\hat{c} \log[T])^2])^2\biggl)\frac{(s-1)}{12}(\hat{c} \log[T])^{s-2}+...
\label{eq:MIMO_asymptotic}
}
This method provides a good approximation \footnote{This is used more as an illustrative example and a better approximation can be found if more care is taken on how to split up the integral which will depend on both $\alpha$ and $\eta$.} provided $s = \frac{\alpha + 2}{\eta} \le 1$and $T > 10$. 

\textit{Example II: Quasi Disk Model}
The quasi-disk model (Sec. \ref{s:Connection_fns})  is a piecewise connection function  which can model a change in channel conditions; for example transitioning from a clutter free environment to a cluttered one. 
In general, assuming a soft-disk model transitioning to a decay function one at $r_-$, through application of eq.\eqref{eq:iso} we obtain, 
\es{&-\frac{\log\mathbb{P}_{\text{iso}}^T(\bxi)}{\bar{N\!c}} = (1- (1-\wp)^T) \int_{\mathcal{W} \cap B_{\bxi}(r_-)}\!\!r^{\alpha+1}g_\phi(\theta)\dd r \dd \theta\\
&+\int_{\mathcal{W} \cap \mathcal{A}_{\bxi}\!(\!r_-\!,\!r_+\!)}\!\!\!\left(\!1\!-\!\left(\!1-\!\wp\!+\!\wp\left(\frac{\!r\!-\!r_-}{r_+\!-\! r_-}\!\right)^{\!\!\!\mu}\!\right)^{\!\!\!T}\!\right)r^{\alpha+1} \dd r g_\phi(\theta) \dd \theta
\label{eq:quasi_gen}}

We can use the previous result for the soft-disk model (see Table \ref{table:table1}) for the first term on the right hand side in eq\eqref{eq:quasi_gen}.
The second term, (denoting the inner radial integral as $I_{\text{radial}}$) can only be given in semi-analytic form using the previously outlined methods when $\bxi \neq 0$.
That is to say we are left with an integral of the form  $\int_0^\phi g_\phi(\theta)(...\!_2 F_1 \left(a, b;c;\xi \cos \theta\right)) \dd \theta$, where $\!_2 F_1 \left(a, b;c;z\right)$ is the Gauss hypergeometric function,  which cannot be computed. 
For simplicity we focus on the case when $\mu = 1$ and let $r_+ = \kappa r_-$.
From method II we need to compute the radial integral 
\es{I_{\text{r}}(\alpha+1)&= \int_{r_-}^{\kappa r_-}\!\!\!\left(\!1\!-\!\left(\!1-\!\wp\!+\!\wp\left(\frac{\!r\!-\!r_-}{r_+\!-\! r_-}\!\right)^{\!\!\!\mu}\!\right)^{\!\!\!T}\!\right)r^{\alpha+1} \dd r 
\label{eq:radial_integral}}

Through direct calculation of the integral in eq.\eqref{eq:radial_integral}, 
\es{I_{\text{r}}(\alpha+1) &= \frac{r_-^{2+\alpha}}{2+\alpha}\biggl(\kappa^{2+\alpha} - 1 + \left(\frac{1}{\Delta(1 - \kappa)}\right)^T\\
&\times\left(\psi\left(T, \alpha, \wp\Delta\right) - \kappa^{2+\alpha}\psi\left(T, \alpha, \kappa\wp\Delta\right) \right)\biggl) 
}
where $\psi\left(T, \alpha, \Delta\right) = \!_2F_1\left(-T;2+\alpha;3+\alpha;\Delta\right)$, $\Delta = \frac{1}{1 - \kappa(1-\wp)}$ and $\kappa \neq \frac{1}{1-\wp}$ so $\Delta \neq 0$.
For the case when $\kappa = \frac{1}{1-\wp}$ we use the following limit, 
\es{\lim_{c\to 0} c^T(-1)^T \!_2F_1\left(-T, a, b, \frac{1}{c}\right) = \frac{\Gamma[b]\Gamma[a+T]}{\Gamma[a]\Gamma[b+T]}}
We now directly use the above result to give $I_{\text{radial}}$ when $\kappa = \frac{1}{1-\wp}$, 
\es{I_{\text{r}}(\alpha\!+\!1\!)\!&= r_-^{2+\alpha}\left(\frac{\kappa^{2+\alpha}\!-\!1}{2+\alpha}\!+\! \frac{\wp^T(1-\kappa^{2+\alpha+T})}{(2+T+\alpha)(\kappa-1)^T}\right)G_\phi}
We can now use method II to provide a general approximation for $I_{\text{radial}}(\alpha+1)$
\es{-\frac{\log\mathbb{P}_{\text{iso}}^T(\xi_x)}{c \bar{N}} &=\!(\!1\!-\!(\!1\!-\wp)\!^{T}\!)\biggl(\frac{r_0^{\alpha+2}}{\alpha+2}G_\phi + F_c r_0^{\alpha+1} \xi_x \biggl)\\
&+ I_{\text{r}}(\alpha+1) G_\phi\! + (\alpha+1)I_{\text{r}}(\alpha) F_c \xi_x}

In the limit as $T\to \infty$ the probability of connection converges to the void probability for the ball of radius $\kappa r_-$. 
\es{\lim_{T\to \infty}\mathbb{P}_{\text{iso}}^T(\bxi) \to V_\mathcal{B}(\bxi,\kappa r_-)}

We remark that the quasi disk can be defined such that it has an exponential decay function and the analysis is very similar to that above, the major difference being that the integral $I_{\text{r}}(\alpha)$ is expressed in terms of Roman harmonic numbers rather than hypergeometric functions.

\begin{table*}
\centering
    \caption{}
    \begin{tabular}{ | l | p{15cm} | }
    \hline
    Model& Approximations for the probability a node $\bxi$ is isolated at time $T$\\ \hline
    SDM  & $\exp\biggl(-\bar{N} c (1-(1-\wp)^T)\biggl(\frac{r_0^{\alpha+2}}{\alpha+2}G_\phi + G_c(\omega) r_0^{\alpha+1} x + r_0^\alpha G_2(\omega) x^2 \biggl) \biggl) $\\ \hline
    SAM & $ \exp\biggl(-\bar{N} c(1 - (1-\wp)^T)\biggl(\frac{(r_+^{\alpha+2} - r_-^{\alpha+2})}{\alpha+2}G_\phi +  G_c(\omega) (r_+^{\alpha+1} - r_-^{\alpha+1}) \xi_x  + G_2(\omega) (r_+^{\alpha} - r_-^{\alpha}) x^2\biggl)\biggl)$  \\ \hline
   QDM  &$\exp\biggl(-\bar{N} c\!(\!1\!-\!(\!1\!-\wp)\!^{T}\!)\biggl(\frac{r_0^{\alpha+2}}{\alpha+2}G_\phi + F_c r_0^{\alpha+1} \xi_x \biggl) + I_{\text{r}}(\alpha+1) G_\phi\! + (\alpha+1)I_{\text{r}}(\alpha) G_c(\omega) x\biggl)$    \\
    \hline
    Rayleigh  & $ \exp\biggl(-\bar{N} c\frac{r_0^{\alpha+2}}{\eta}\Gamma\left[\frac{2+\alpha}{\eta}\right]H_{T,1}^{\frac{2+\alpha}{\eta}}G_\phi  +  r_0^{\alpha+1} \Gamma\left[\frac{1+\alpha}{\eta} + 1\right]H_{T,1}^{\frac{1+\alpha}{\eta}}x G_c(\omega) +...\biggl)$ \\ \hline
    Waxman  & $ \exp\biggl(-\bar{N} cr_0^{2+\alpha}\Gamma[2+\alpha]\bar{H}^{2+\alpha}_{T,\beta}G_\phi +  r_0^{\alpha}\Gamma\left[\alpha+1\right]\bar{H}^{1+\alpha}_{T,\beta} G_c(\omega) x + ... $  \\ \hline
    Interference  & $\exp\biggl(-\bar{N} c\!\frac{H_T}{(2+\alpha)c_{\mathcal{I}}} +\!\frac{(c \bar{N})^{\frac{1}{2+\alpha}}}{(c_{\mathcal{I}} G_\phi)^{\frac{1+\alpha}{2+\alpha}}}\Gamma\left[\!\frac{3+2\alpha}{2+\alpha}\!\right]H_T^{\frac{1+\alpha}{2+\alpha}} x G_c(\omega) + O(x^2)\biggl)$ \\
    \hline
    MIMO  & $ \exp\biggl(-\bar{N} c\frac{r_0^{\alpha+2}G_\phi}{\eta} I_1\left(\frac{\alpha+2}{\eta}\right) + \frac{(\alpha +1)r_0^{\alpha+1} G_c \xi_x}{\eta} I_1\left(\frac{\alpha+1}{\eta}\right)\biggl)$  \\
    \hline
    \end{tabular}
    \vspace{0.35cm}\label{table:table1}
    
Table of approximations for a range of different connection functions $\h$ calculated form eq.\eqref{eq:iso}, see Table \ref{table:table_H} for definition of connection functions and symbols. 
    Recall that $\bxi = (\xi_x, \xi_y)$ is in Cartesian coordinates while $\bxi = (x, \omega)$ is in polar coordinates.
    Refer to Section \ref{s:Connection_fns} for definition and explanation of parameters used. 
  \end{table*}

\section{Numerics}\label{s:numerics}

\begin{figure*}
\centering
\includegraphics[scale = 0.3]{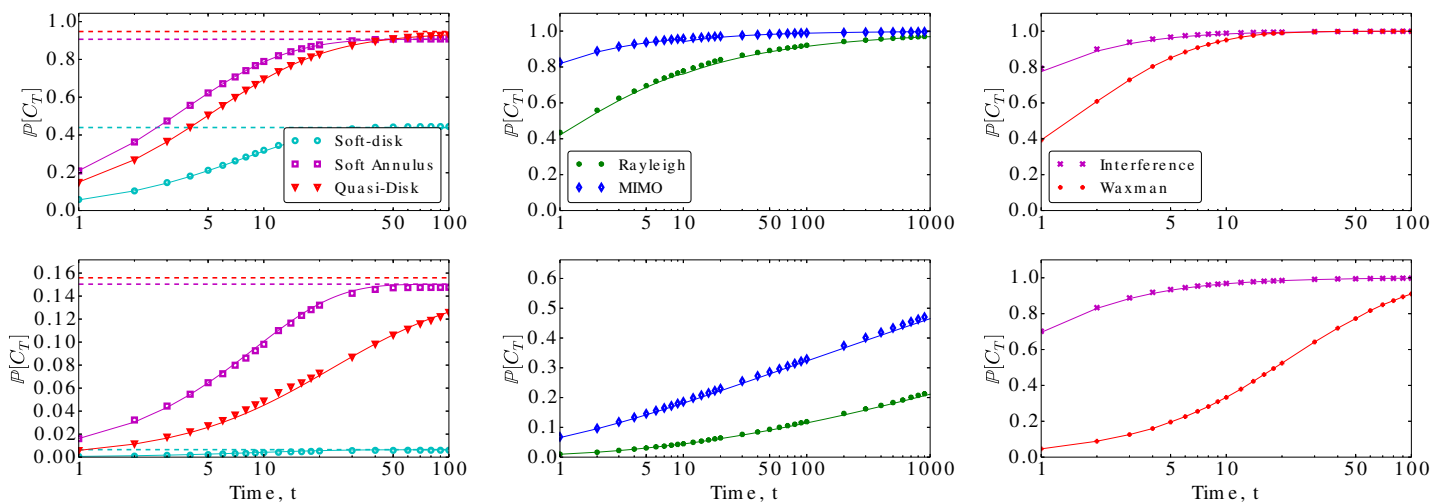}
\caption{The probability that a node located near the corner at $\bxi = (0.2, \phi/2)$ is connected for different connection functions.
The top panel and bottom panels have parameter $\alpha = 0,2$ respectively, comparing the impact the spatial distribution of nodes in the network has on connectivity.
The dashed lines represent the void probabilities, the solid thin  lines are the approximations (given in Table \ref{table:table1}) and the circle markers are simulated points. 
For the SDM and SAM cases  the approximations are found form translating the densities (Sec. \ref{SD-section}); the Rayleigh, Waxman and Interference case use Method I (Sec \ref{s:methodI}) and the relevant asymptotic approximations, whilst the MIMO case uses method II (Sec \ref{s:methodII}).
Parameters $\wp = 0.1, \phi = \frac{\pi}{4}, \beta = 0.5, r_0 = 1.0; r_- = 1.0, r_+ = 2.5, L = 10$ and $\eta =4,4,6$ for the Rayleigh, MIMO and Interference cases respectively.}
\label{fig:fig_Iso}
\end{figure*}

\begin{figure*}
\centering
\includegraphics[scale=0.30]{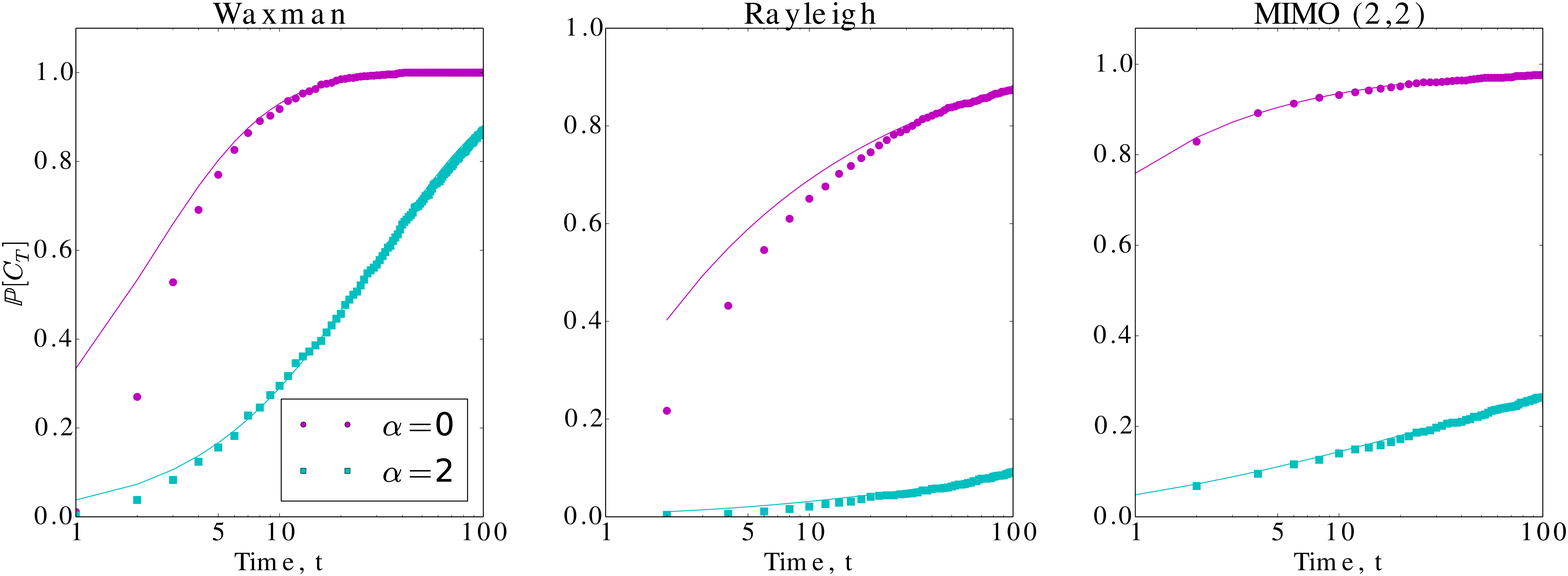}%{"Fig1_approx_asym_vs_no_iso_inf_wedge_epsx_0_epsy_0phi_0_25pieta_2_beta_0_5_r0_1__L_10"}
\caption{The probability that a node is connected at time $T$ located at the corner compared with a simulation of the probability every node in the network is connected at time $T$; clearly the node near the corner is the last to connect.
Parameters used: $L = 10$, $\phi = \pi/4$, $\eta = 2$, $\beta  = 0.5$, $\bxi = (0.2,\phi/2)$ and $r_0 = 1$.   }
\label{fig:No_iso}
\end{figure*}

Firstly, the approximations provided in the previous sections, included in Table  \ref{table:table1}, are a good fit for the simulated data points, see Fig\ref{fig:fig_Iso}.
One general observation (all connection functions expect for the interference case) is that the probability of connection tends to its maximum much faster for larger $\bar{N}$ (similarly for larger $r_0$ or smaller $\alpha$) as the local neighbourhood becomes increasingly dense. 
For the interference model the change in connection probability is much smaller as the density changes since only the second term depends on $\bar{N}$, a result of the trade-off between connectivity and interference, and as nodes are added to the network the probability of connections are counter balanced by the increase in interference field.

\subsection{Connection functions with compact support}
For connection functions with compact support the probability $\bxi$ is connected tends to the  complement of the void probability and is represented by the dashed lines in  Fig\ref{fig:fig_Iso}.
That is to say, the limiting behaviour is restricted to there existing a node within the connection range, i.e the void probability which is characterised by the PPP $\Phi$ and $r_0, r_-,r_+$. 
Such connection functions are employed in the modelling of wireless sensor networks, and an easy way to ensure connectivity  is to enforce an underlying structure to the network (lattice) so that the maximum distance between any two sensors is at most $r_0$.
However, in dense networks (or equivalently when the typical connection range is large)  where devices are located predominately within the bulk, it is likely that a lattice structure is not needed and will only waste resources. 
Our results highlight how the boundaries, along with inhomogeneities, significantly decrease the connection probability.
For example, if $r_0 = 1; L = 10; \phi = \pi/2$ then the mean degree when $\alpha = 0$ is $ \approx 0.407$ compared with    $\approx 0.003$ when $\alpha = 2$. 
As a result, in networks that exhibit such behaviour it is likely nodes need only need be added near the boundary to ensure connectivity. 
\subsection{Connection functions with global support}
For connection functions with infinite support we see that the probability approaches the complement of the probability the PPP is non-empty, see section \ref{s:methodI}, and does so faster for a larger $r_0, \bar{N}$ and smaller $\alpha$. 
This behaviour is a finite domain effect, and if we condition on there being at least one point in the PPP (or else use a Binomial Point Process), then 
$\mathbb{P}[C_T] \to^{T\to\infty}  1$ . 
For both the Rayleigh and MIMO cases the asymptotic expansions work well for large $T$, and improve when the path-loss exponent $\eta$ increases (the signal decays faster), or the distribution of points becomes more uniform. 
For the MIMO case a better approximation can be provided for specific $\alpha$, $\eta$ but it is unclear how to improve it for the general case. 
However, as the probability for long links increases, such as in the Waxman case, the usefulness of the large $T$ approximation is limited to the uniform case, but for the non-uniform case the approximation for very inhomogeneous networks works well.  

The connectivity of infinite networks are obstructed by corner nodes, provided some assumptions on the density that it grows away from the corner, $\alpha>0$.
If however, the PP is uniform, or even if $\alpha < 0$ then  the network may never connect, you may have infinitely many isolated nodes \cite{dettmann2017isolation}.

\subsection{Full connectivity}
In static networks $P_{fc}$ is defined as there existing a multi-hop path between any two nodes in the network. 
In a temporal network this is more complicated since there exists a network with directional (causal) paths between nodes.
We introduce a weaker sense of full connectivity, that is the probability that every node in the network has made at least one link to some other at, or prior to, time slot $T$; we will denote this as $P_{fc}^T$.    
Analogous to other work, we want to make use of there being no isolated nodes to approximate that of $P_{fc}^T$. 
Indeed, focusing on the idea that boundary nodes are likely to be "more isolated" we see that nodes near the corner are the last to connect when links are independent,see Fig\ref{fig:No_iso}. 
Naturally, when considering interference this behaviour is not necessarily true since nodes near the bulk may be in outage if the interference field is to high; in fact the boundary may help connectivity due to a decreased interference field. 
Essentially, we have shown in Fig\ref{fig:No_iso} that the time for every node in the network to form a link is determined by how long the highly isolated nodes take to form a link. 
Furthermore, provided the network is dense enough and $\alpha \gtrapprox 1$ then it is likely the first causal path occurs from any node in the network to a boundary node when the boundary node makes a single connection.  

For infinite networks with non-uniform measure isolated nodes are likely be play a more significant role for $P_{fc}$ \cite{dettmann2017isolation}. 
For example, if $\alpha  \le 0$ then the the number of isolated nodes is infinite, and thus $P_{fc} = 0$ can never be achieved, whereas when $\alpha > 0$ the behaviour is likely to be determined by highly isolated nodes \cite{dettmann2017isolation}.

\section{Conclusions}\label{s:conclusion}

In this paper we look at the impact local geometries and non-uniform densities have on wireless networks, and show those nodes near the corners dominate the global connectivity properties of the network, especially when the local neighbourhood is sparse. 
The location of nodes were modelled by a non-uniform Poisson Point Process in a triangular domain, and links were formed during each time slot, independently from the past, based on a probabilistic connection function that depended on node separation. 
The time for information to flow through the network was assumed to be much less than the time scale for mobility, thus we could assume the location of nodes to be fixed (albeit not uniformly distributed).
More specifically, two methods were provided for calculating the probability a node near the corner was isolated at time $T$ for a general connection function, where some examples were given from the wireless literature. 
The first method was used to generate closed form expressions for general densities $r^\alpha g_\phi(\theta)$ (not necessarily vanishing at the borders) which required calculating the higher moments of the connection function.
For more complicated connection functions (where the higher order moments were not integrable, easy to calculate or did not provide closed form expressions) a second method was proposed in order to reduce the number of integrals that needed to be computed.
The latter required that the density went to zero along one of the boundaries, which was not a requirement in the first method. 
For all the connection functions discussed in this paper we provide asymptotic approximations for large $T$ and/or large $\alpha$ and show they are a good fit compared with simulations.
Furthermore, we also provided an approximation for full connectivity that it is those nodes near the corner (and with few close neighbours) that are highly isolated that are the main obstacle.
This naturally assumes that nodes within the bulk have made multiple connections in the previous time slots and thus any information has flowed through the rest of the network.  

Although the examples given in this paper are from the wireless literature they can easily be extended to different connection functions.

This work can provide insight into the demand for future network design, for example smaller access points (small base stations as those required for the deployment of 5G networks) should be deployed near boundaries or regions of low density to  ensure connectivity. 
In general, the properties of isolation near the boundaries in networks with non-uniform measure could in theory be exploited to halt the spread of forest fires, or disease, where border nodes represent a bridge between networks with high betweenness centrality.

In this paper we assume a static distribution of nodes but it would be interesting if the locations of receiver nodes vary with time according to some mobility model and how this impacts on the global connectivity of the network.
Future work could also include finding an effective method to approximate the number of connected subgraphs in finite networks with non-uniform measure. 
This would allow a deeper understanding into the transition from disconnected to a fully connected network. 
A closed form expression involving multiple integrals can be derived using the theory of point processes but approximating the expected number of clusters of a particular size remains open; even for the case of isolated nodes  with non-uniform measure.
This is of particular interest since it has been recently shown that the property of disconnection beings more heavily coupled with that of isolated nodes when the density is non-uniform \cite{dettmann2017isolation}.

\section{Acknowledgements}
The authors would like to thank the directors of the Toshiba
Telecommunications Research Laboratory for their support.
This work was supported by the EPSRC [grant number
EP/N002458/1]. In addition, Pete Pratt is partially supported
by an EPSRC Doctoral Training Account.

\bibliographystyle{ieeetr}
\bibliography{Ref}	
\end{document}